\documentclass[fleqn,12pt,twoside]{article}
\usepackage{espcrc1}

\usepackage{graphicx}
\usepackage[figuresright]{rotating}

\usepackage{placeins}

\newcommand{\ud}{\mathrm{d}}

\title{Baryon and antibaryon production in hadron-hadron and 
       hadron-nucleus interactions }

\author{M.Kreps\address[MCSD]{ Institute of Physics,
Comenius University,
                              Mlynsk\'a dolina, 842 48
                              Bratislava, Slovakia}
 for the NA49 Collaboration\footnote{For full author 
list see M.~van~Leeuwen's contribution to these proceedings}}
\begin{document}

\maketitle

\begin{abstract}

Cascade baryon and anti-baryon yields have been measured in
$p+p$
and $p+A$ collisions. After extraction of the projectile
component in p+A interactions close similarities with $A+A$ collisions
concerning the nuclear enhancement factors are observed. In addition
the importance of effects related to projectile isospin and to net
baryon stopping is pointed out.
\end{abstract}

\section{Introduction}

The NA49 experiment is unique in measuring $\Xi^-$ and
$\overline\Xi^+$ production
over the full range of hadronic interactions, from $p+p$ collisions via
centrality controlled $p+Pb$ to central $Pb+Pb$ reactions. This offers
interesting possibilities of comparison, in particular concerning the
nuclear enhancement factors in $p+A$ and $A+A$ collisions. The extraction
of these factors is straightforward for the symmetric $p+p$
and $Pb+Pb$ 
interactions, however the interpretation of central rapidity yields
from $p+A$
collision poses certain problems of normalization or, rather, separation
of projectile and target contributions. In addition, the effects of projectile
and target isospin and of net baryon stopping have to be properly addressed.

\section{A two component picture of hadronic interactions}

\begin{figure}[t]
\centering
\vspace{-5pt}
\includegraphics[width=11cm]{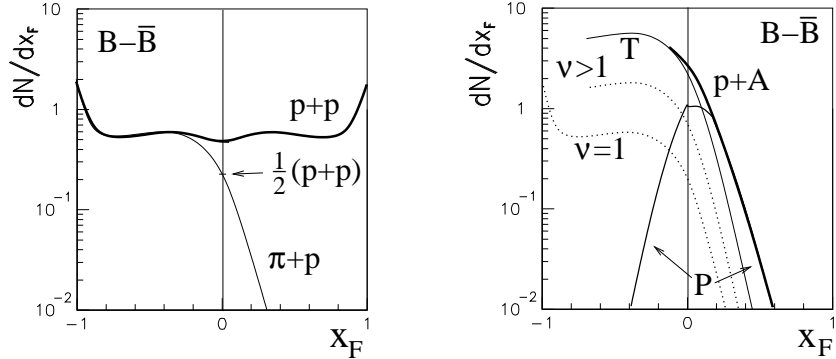}
\vspace{-25pt}
\caption{Hypothetical comparison of net baryon spectra in
$pp$, $\pi p$, $pPb$ and $\pi Pb$ reactions. Spectra for
$pp$, $pA$ are splitted to target and projectile component.}
\label{fig:two-comp}
\end{figure}

Invoking the absence of charge and flavor exchange at SPS energies one
may postulate hadronic factorization in the sense of splitting particle
production into two independent components, one from the target and one
from the projectile, fig.\ref{fig:two-comp}.
This picture has been verified \cite{art:fischer-this} for net
proton production both for $p+p$ and $p+Pb$ interactions, using pion beams.
It has been shown that particle densities 
in $p+Pb$ collisions 
at central rapidity are built
up from a~target part which is proportional to the number of projectile
collisions, $\nu$, and a~projectile part which carries the imprint of the
hadron which has undergone $\nu$ collisions.
The presence of nuclear enhancement in $p+A$ collisions can accordingly be
analyzed along two different lines: either this enhancement
$E$ is equally
distributed to target and projectile components ("wounded nucleon" model),
%
%
\begin{equation}
\left(\frac{\ud n}{\ud X_F}\right)^{p+A}_{X_F=0}\,=\,
\left[\frac{\nu\cdot \alpha}{2}\cdot E(\nu)\,+\,
\frac{1}{2}\cdot
E(\nu)\right]\cdot\left(\frac{\ud n}{\ud
X_F}\right)^{p+p}_{X_F=0}
\label{eq:pA-wound}
\end{equation}	    
or it is fully credited to the projectile contribution,
%
%
\begin{equation}
\left(\frac{\ud n}{\ud X_F}\right)^{p+A}_{X_F=0}\,=\,
\left[\frac{\nu\cdot \alpha}{2}\,+\,
\frac{1}{2}\cdot
E(\nu)\right]\cdot\left(\frac{\ud n}{\ud
X_F}\right)^{p+p}_{X_F=0}
\label{eq:pA}
\end{equation}	    
The latter possibility has been shown to be valid by NA49 at least for
net protons \cite{art:fischer-this}. Alpha is an isospin 
factor as discussed below.

\section{Cascade enhancements}
\begin{figure}
\centering
\includegraphics[width=6cm]{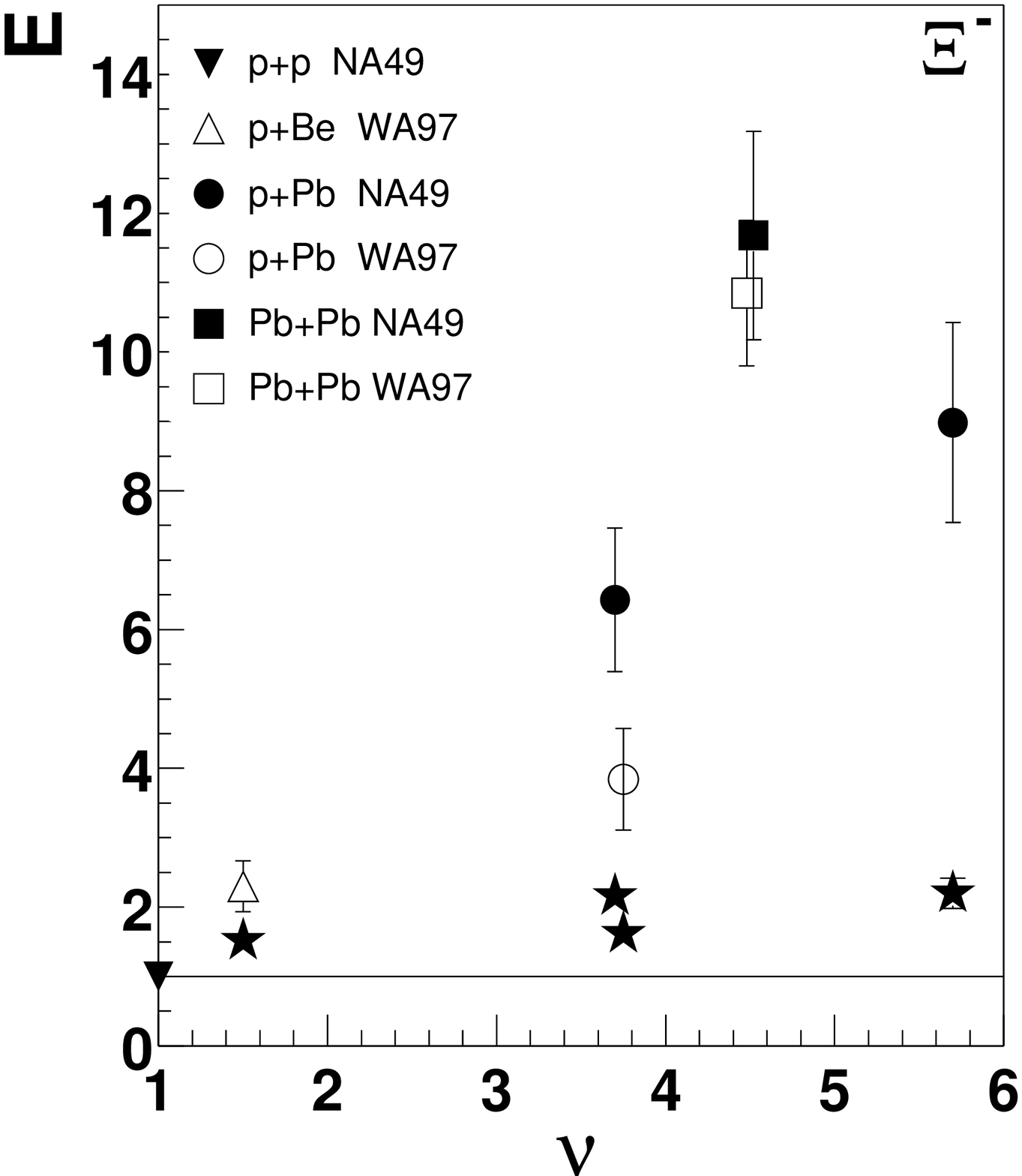}
\hspace{1cm}
\includegraphics[width=6cm]{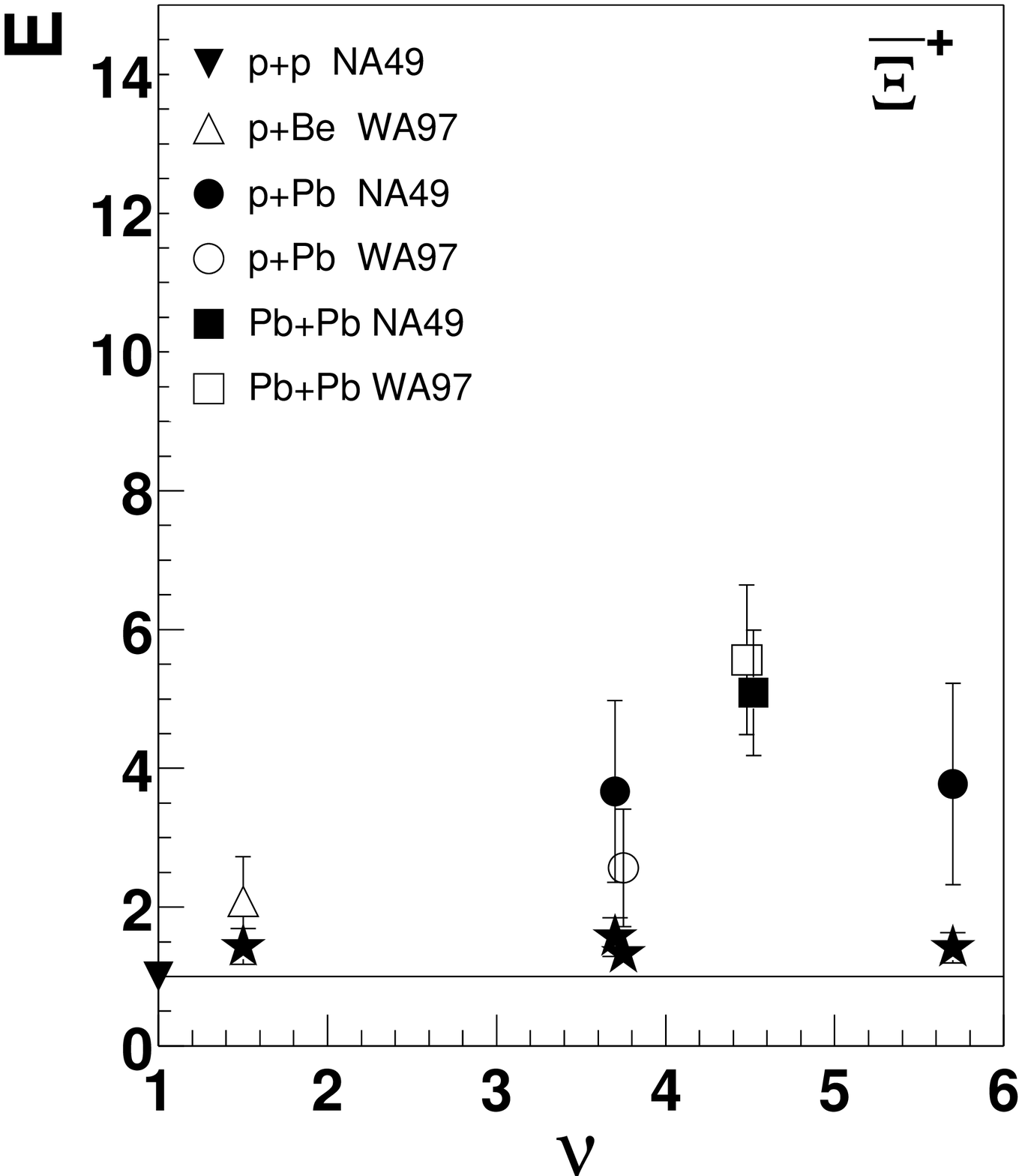}
\caption{Enhancement factor for $\Xi^-$ and $\overline\Xi^+$
at midrapidity calculated with two different assumptions
(see text). The stars indicate enhancement factors in
$p+A$ reactions calculated from equation \ref{eq:pA-wound}}
\label{fig:enhancement}
\end{figure}
\begin{figure}
\centering
\includegraphics[width=6cm]{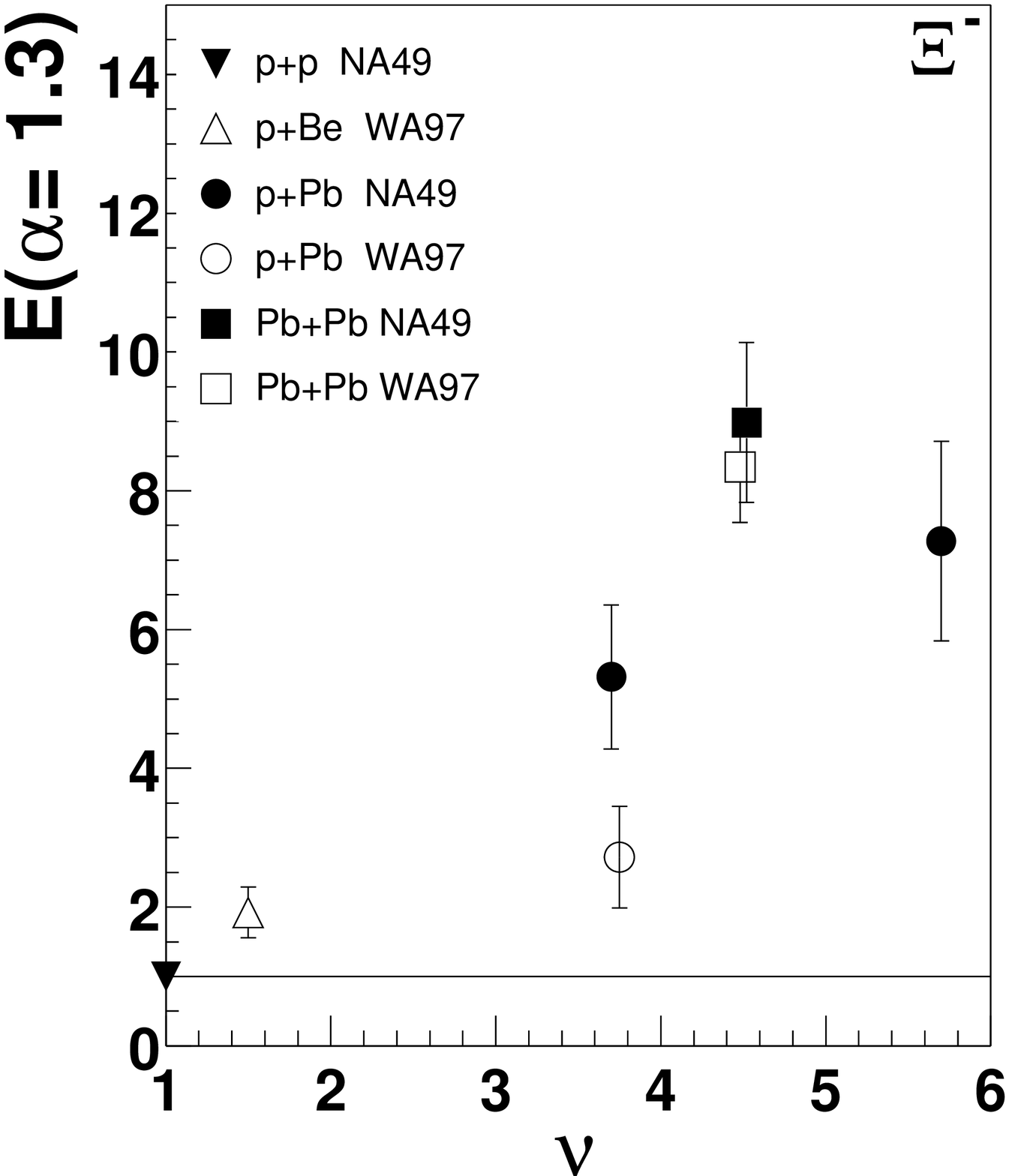}
\hspace{1cm}
\includegraphics[width=6cm]{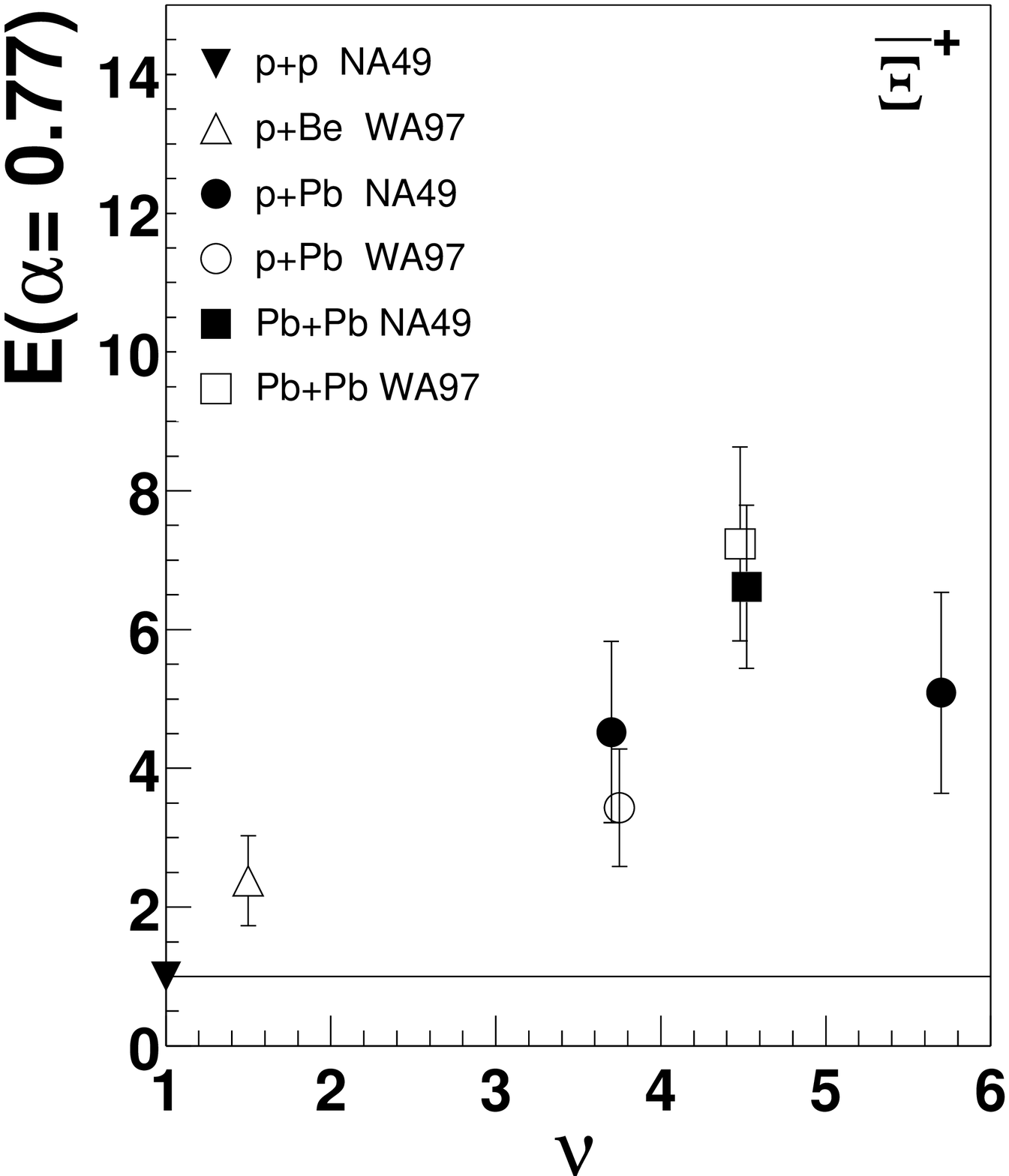}
\caption{Isospin corrected enhancement factor for $\Xi^-$ 
and $\overline\Xi^+$ at midrapidity calculated from equation
\ref{eq:pA}.}
\label{fig:enhancement-iso}
\end{figure}

Using the measured $\Xi$ and $\overline\Xi$ cross sections
\cite{art:tanjaqm,art:wa97}  
the extracted
enhancement factors turn out to be largely different following the two
prescriptions, as shown in fig.\ref{fig:enhancement} for $\alpha=1$.
Clearly, the projectile
enhancement obtained from method (\ref{eq:pA}) in $p+A$ collisions 
is very similar to
the one seen in $Pb+Pb$ reactions whereas method
(\ref{eq:pA-wound}), due to the overriding
target contribution, yields little or no overall enhancement
in $p+A$ collisions.
A noteworthy feature of fig.\ref{fig:enhancement} is the sizeable 
difference in enhancement between
$\Xi^-$ and $\overline\Xi^+$ which amounts to about a factor of two. 

\section{Isospin effects}

Since heavy nuclei contain about $60\%$ neutrons, eventual differences 
in particle
yields between neutron and proton fragmentation have to be carefully 
taken into
account. It has been shown
\cite{art:fischer-this,art:fischer} that for anti-protons for
example there is a $50\%$
yield difference between neutron and proton induced
elementary interactions. 
This difference
has been traced to an important component of asymmetric baryon pair production
(e.g. $n\overline{p}$ or $p\overline{n}$) which strongly depends on 
projectile isospin.
In the absence of measurements of $\Xi$ production from neutrons, 
it seems reasonable
to assume that similar pair production mechanisms hold for
$\Xi\,\overline\Xi$ production.
Given the charge structure of the isospin triplet of
$\Xi\,\overline\Xi$ states,
\begin{displaymath}
\Xi^-\overline\Xi^0 \quad
\Xi^-\overline\Xi^+/\Xi^0\overline\Xi^0 \quad
\Xi^0\overline\Xi^+
\end{displaymath}		 
a decrease in $\overline\Xi^+$ and an increase of $\Xi^-$ production is 
to be expected when 
switching from proton to neutron projectiles. Taking account
of the $40/60\%$ mixture
of protons and neutrons in $Pb$ nuclei and assuming the same asymmetry 
factor as measured for $S=0$ baryons this results in an effective 
isospin factor $\alpha=1.3$
for $\Xi^-$ and $0.77$ for $\overline\Xi^+$. The isospin corrected 
enhancement factors (see equation \ref{eq:pA}) are shown
in fig.\ref{fig:enhancement-iso}. 
Apparently the similarity between $p+A$ and $Pb+Pb$ interactions is preserved
and the large asymmetry between $\Xi^-$ and $\overline\Xi^+$ enhancements, 
fig.\ref{fig:enhancement}, is strongly reduced.

\section{Net baryon stopping}

\begin{figure}[t]
\begin{minipage}[t]{.45\textwidth}
\centering
\vspace{-15pt}
\includegraphics[width=6cm]{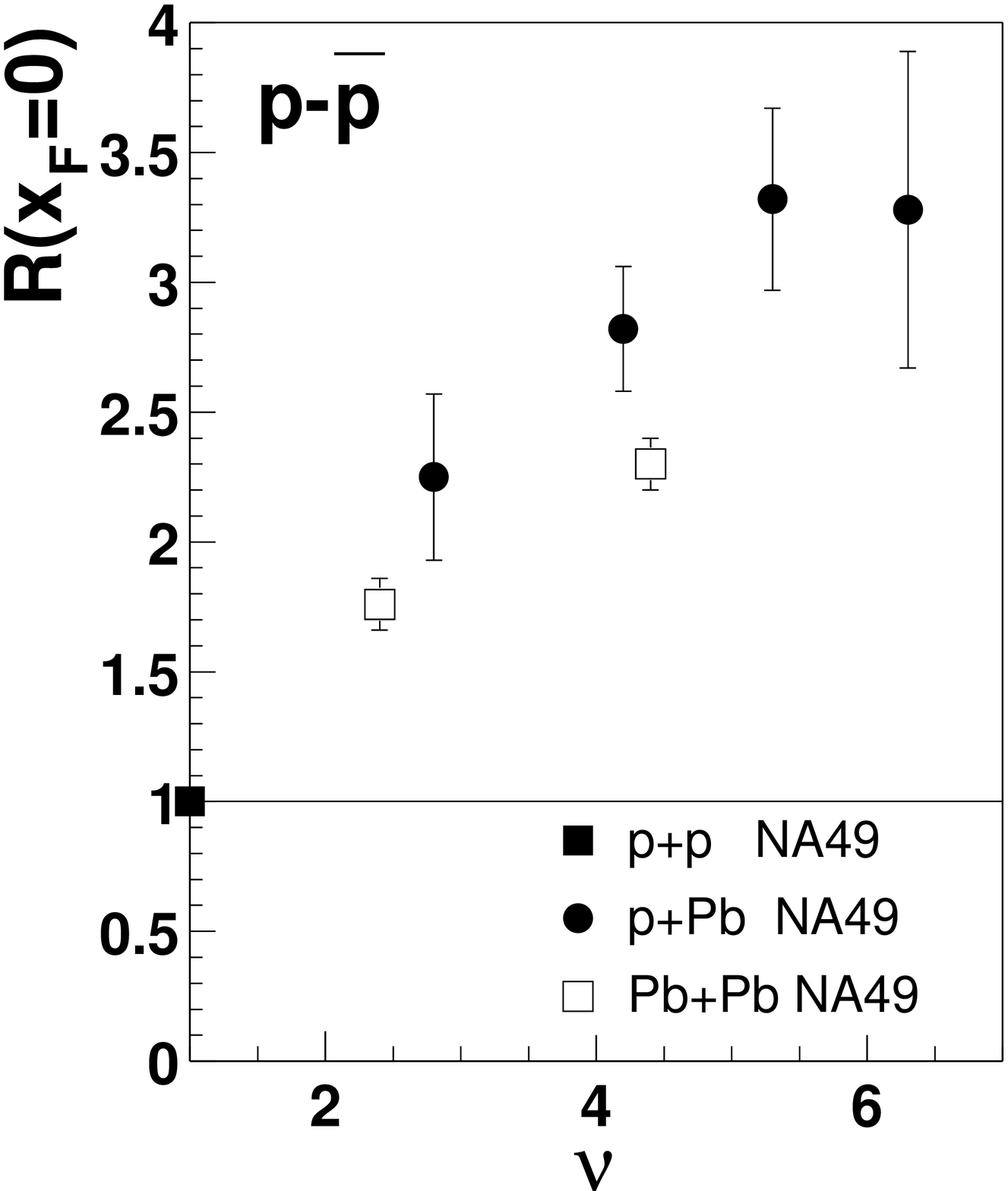}
\vspace{-15pt}
\caption{Ratio of projectile component of net protons in
$p+A$ reactions to projectile component of net protons in
$p+p$. For details see \cite{art:fischer-this}.
}
\label{fig:netprot}
\end{minipage}
\hspace{.05\textwidth}
\begin{minipage}[t]{.45\textwidth}
\centering
\vspace{-15pt}
\includegraphics[width=6cm]{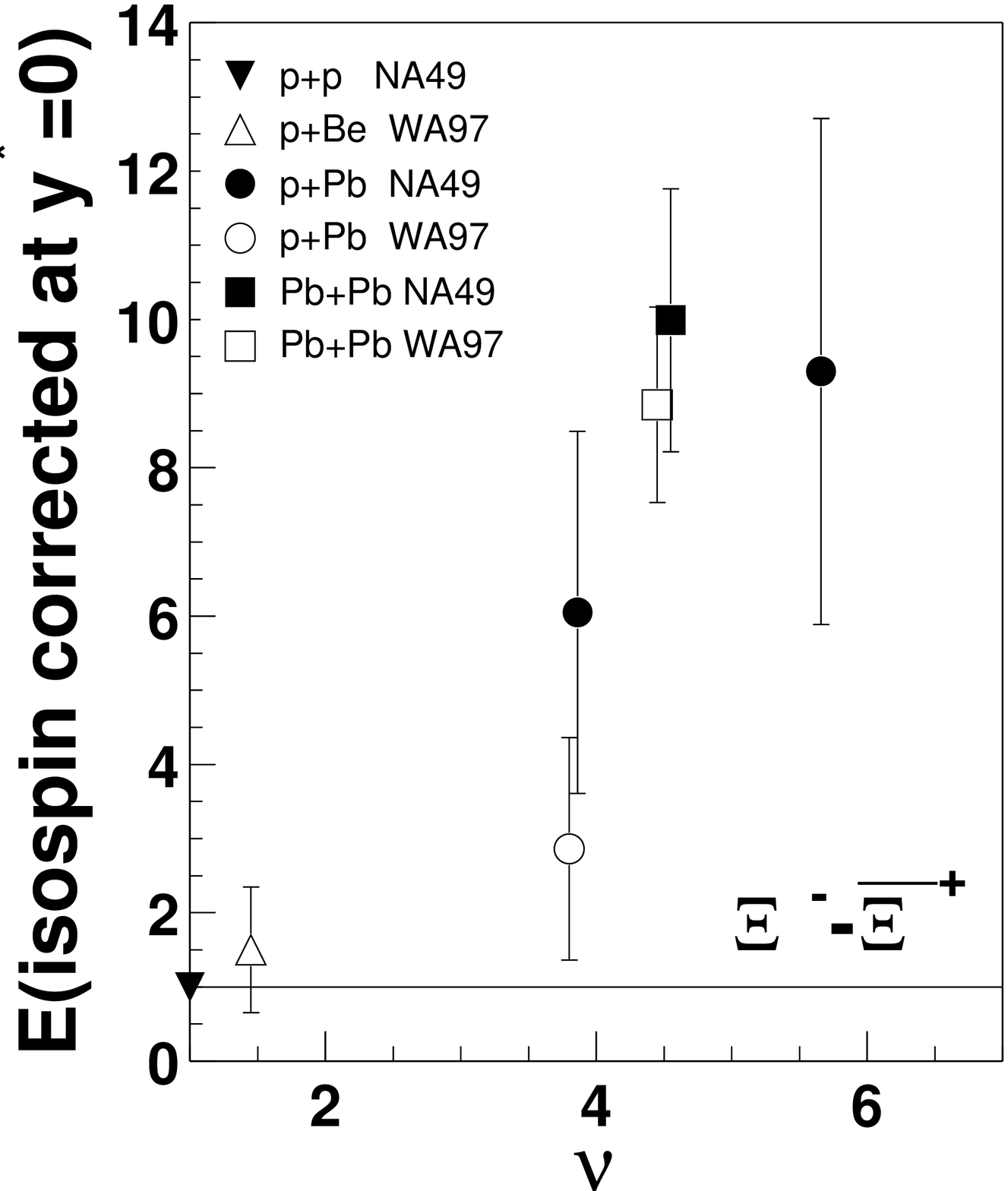}
\vspace{-15pt}
\caption{Isospin corrected enhancement factor for net
$\Xi^-$.
}
\label{fig:netxi}
\end{minipage}
\vspace{-10pt}
\end{figure}

The transfer of net baryon number from forward to central region in multiple
collision processes ("stopping") may play an important role in the 
interpretation of central baryon densities. The net proton 
yield from the projectile e.g. increases by a factor of about $3$
at central rapidity 
when comparing $p+p$ and central $p+Pb$ interactions
\cite{art:fischer-this} (see fig. \ref{fig:netprot}).
For $\Xi^-$ production, corresponding studies would imply the 
measurement of complete $x_F$ distributions in pion and proton 
induced collisions. Such measurements are not available. 
Nevertheless one may conjecture that --- given the steeper
$x_F$ dependence of $\Xi^-$ as compared to $p$ --- the 
"enhancement" of net $\Xi^-$ at $x_F=0$ in multiple collision
processes should be smaller. The experimental situation 
concerning net $\Xi^-$ production
is given in fig.\ref{fig:netxi} which shows the enhancement factors as 
function of $\nu$. 
One may conclude that --- compared to
$\overline\Xi^+$ --- the relative increase is of order $1.5$ which
might well originate from net $\Xi^-$ stopping.
%



\begin{thebibliography}{9}
\bibitem {art:fischer-this}{H.~G.~Fischer, these
proceedings.}
\bibitem {art:tanjaqm}{T.~\v Su\v sa for the NA49
Collaboration, Nucl.\ Phys.\ {\bf A698}(2002) 491c.}
\bibitem {art:wa97}{F.~Antinori {\it et al.} [WA97
Collaboration], Nucl.\ Phys.\ {\bf A661}(1999) 130c.}
\bibitem {art:fischer}{H.~G.~Fischer for the NA49 
Collaboration, Acta\ Phys.\ Pol.\ \textbf{B33}(2002) 1473.}
\end{thebibliography}
\end{document}